\documentclass[conference,10pt]{IEEEtran}
\usepackage{epsfig,rotating,setspace,latexsym,amsmath,epsf,amssymb,amsfonts,bm,theorem,epstopdf}
\usepackage{cite,authblk}
\usepackage{bbm}
\usepackage{color}
\usepackage{mathtools}
\usepackage{algorithm}
\usepackage[noend]{algpseudocode}
\usepackage{comment}

\algrenewcommand\algorithmicforall{\textbf{foreach}}
\algrenewcommand\algorithmicindent{.8em}

\IEEEoverridecommandlockouts
\allowdisplaybreaks

\newcommand{\figref}[1]{\figurename~\ref{#1}}

\title{Dynamical Dorfman Testing with Quarantine}
	
\author{Mustafa Doger \qquad Sennur Ulukus\\
\normalsize Department of Electrical and Computer Engineering\\
\normalsize University of Maryland, College Park, MD 20742\\
\normalsize  \emph{doger@umd.edu} \qquad \emph{ulukus@umd.edu}}

\begin{document}

\maketitle

\begin{abstract}
    We consider dynamical group testing problem with a community structure. With a discrete-time SIR (susceptible, infectious, recovered) model, we use Dorfman's two-step group testing approach to identify infections, and step in whenever necessary to inhibit infection spread via quarantines. We analyze the trade-off between quarantine and test costs as well as disease spread. For the special dynamical i.i.d.~model, we show that the optimal first stage Dorfman group size differs in dynamic and static cases. We compare the performance of the proposed dynamic two-stage Dorfman testing with state-of-the-art non-adaptive group testing method in dynamic settings. 
\end{abstract}

\section{Introduction}
Group testing, introduced by Dorfman \cite{Dorfman1943} in 1943, is a powerful tool to identify infected individuals in a population using a minimum number of tests. In group testing, samples from multiple individuals are mixed  and tested together. If disease prevalence in the population is low, mixed samples are not contaminated with high probability, which in turn results in gains in terms of number of tests. 

Group testing algorithms where tests to be performed are designed in advance are called non-adaptive, and algorithms that use the results of previous tests are called adaptive \cite{gt-survey}. Traditional group testing approaches assume that infections are either i.i.d.~in a population (i.i.d.~model) or the number of infections is known (combinatorial model) \cite{hwang-book, iid-model, gt-survey}. Both of these assumptions are naive and do not hold in reality. Sparked by the ongoing covid-19 pandemic, many recent works have considered infection models with a community structure and proposed algorithms that exploit the community structure in order to reduce the required number of tests \cite{diggavi-community, ozgur-community, arasli-community, diggavi-community2}. 

Another limitation of traditional group testing approaches is the assumption that the infection status of individuals is static, whereas in practice there is a time dimension to disease spread. Using the SIR model \cite{sir-model-book}, recent references \cite{diggavi-dynamic-entropy-reduction, diggavi2021dynamic} consider community-aware group testing for the dynamic case. In \cite{diggavi-dynamic-entropy-reduction}, SIR model is based on a continuous-time Markovian process and the aim is to estimate the states of individuals while reducing the number of tests spent on each day using an entropy reduction approach. Entropy reduction refers to the idea that, in order to reduce the number of tests, entropy of each testing stage should be maximized \cite{jaggi-indep, doger-ulukus-noniid}. In \cite{diggavi2021dynamic}, SIR model with a stochastic block model (SBM) is discretized and a theoretical guarantee for the order-optimality of the number of tests spent per day is given. In this paper, we consider a similar discrete-time SIR model with some variations in the community structure, introduce new ideas related to dynamic group testing and analyze their implications.

\subsection{Related Works}
Our work is closely related to Dorfman's group testing \cite{Dorfman1943} and the group testing problem studied in \cite{diggavi2021dynamic} with discrete-time SIR-SBM. In \cite{Dorfman1943}, for an i.i.d.~infection model, Dorfman divides the population of size $N$ into groups of size $s$ and tests these groups. This is the first stage of testing. In the second stage of testing, individuals belonging to contaminated groups are tested individually to find all infections. Dorfman finds the optimum group size $s^*$ to minimize expected number of tests.

In \cite{diggavi2021dynamic}, the population consists of communities, such that the probability of infection spread within a community is larger than the probability of infection spread across communities. With a discrete-time SIR model, infections, infection spread, recoveries and interventions occur only at discrete time instants (days). In addition, test results come with one day delay; the result of a  test registered on day $d$ becomes available on day $d+1$. Based on these assumptions, \cite{diggavi2021dynamic} shows that, given the test results of the previous day, the discrete-time SIR-SBM reduces to a static group testing problem on the current day. As a result, \cite{diggavi2021dynamic} applies existing results from non-adaptive group testing \cite{jaggi-indep} and shows them to be order optimal.

\subsection{Our Contributions}
We adapt Dorfman testing to discrete-time SIR-SBM introduced in \cite{diggavi2021dynamic}. We demonstrate the benefits of using Dorfman's adaptive method compared to the non-adaptive method used in \cite{diggavi2021dynamic}. As a stand-alone problem, we study the effects of quarantining individuals in a positive first stage test in a Dorfman type two-stage adaptive method, and find the optimum group size considering the opposing goals of minimizing the number of tests versus not quarantining uninfected individuals unnecessarily. The optimal group size so obtained is different than the optimal Dorfman group size. 

We then apply the quarantine idea to discrete-time SIR-SBM. We analyze the trade-off between quarantine costs, test costs, and quarantine's effect on infection spread over time. For a special case of dynamic i.i.d.~infection model, we find the optimal dynamic Dorfman group size analytically, and show that it differs from the static Dorfman group size. 

We provide extensive numerical results: For the SIR-SBM model, we compare the performance of the dynamic Dorfman testing proposed here with the state-of-the-art non-adaptive algorithm proposed in \cite{diggavi2021dynamic}. For the i.i.d.~dynamic setting, we show gains in terms of number of tests for static Dorfman group size versus dynamic Dorfman group size.

\section{SIR-SBM Model and Dorfman Adaptation}\label{SIRSBM-section}

\subsection{Discrete-time SIR-SBM}
We consider a population of size $N$, that is partitioned into communities of size $C$ each. On any day $d \in \mathbb{N}$, each individual $i$ belonging to community $j$ is in one of three states: $X_{i,j}^{(d)} \in \{\mathcal{S},\mathcal{I},\mathcal{R}\}$ where $\mathcal{S}$ stands for susceptible, $\mathcal{I}$ for infected and $\mathcal{R}$ for recovered. We assume $\mathcal{R}$ is an absorbing state, whereas states $\mathcal{S}$ and $\mathcal{I}$ are transient. During a day, the state of individuals may transition from $\mathcal{S}$ to $\mathcal{I}$, or from $\mathcal{I}$ to $\mathcal{R}$ according to SBM that is detailed below. At time\footnote{In this paper, we use notations $d^-$, $d^+$, $d^{++}$ to mean the following. Ony any day, we have a fixed reference time, such as 9am. On day $d$, $d^-$ means slightly before the reference time, e.g., 8:59am, when we receive the previous day’s test results; $d^+$ means slightly after the reference time, e.g., 9:01am, when we take measurements and submit (register) tests; and $d^{++}$ means the time between the submission of the tests on day $d$ until the test results are received on day $(d+1)$, i.e., the time duration after $d^+$ and before $(d+1)^-$.} $d=0^{++}$, infection starts to spread among population where each individual has i.i.d.~infection probability, $p_{init}$. On following days, infection spreads according to SBM with parameters $(N,C,q_{1},q_{2},p_{init})$. Namely: If an individual is in state $\mathcal{I}$ at $d^{+}$, then during $d^{++}$ it spreads infection to any of its susceptible neighbors in the same community with probability $q_{1}$, and to members of other communities with probability $q_{2}$. Any individual who was infected at time $d^{+}$, can recover during $d^{++}$ with probability $r$. The discrete-time SIR-SBM captures the disease progression of continuous-time SIR-SBM well \cite{diggavi2021dynamic}. Although we assume that discrete-time instances are days, the model can be discretized with other time-units.

\subsection{Test Results}
Infection status of an individual is denoted by $U_{i,j}^{(d)}$, where $U_{i,j}^{(d)}=1$ iff $X_{i,j}^{(d)}=\mathcal{I}$. A group test $T$ takes samples from a set of individuals $\mathcal{G}_{T}$ and tests them at once. The result of the test is $y_{\mathcal{G}_{T}}=\bigvee_{i \in \mathcal{G}_{T}} U_{i,j}^{(d)}$, where $\bigvee$ stands for binary OR operation. The tests are registered on each day at time $d^{+}$, shortly after that, during $d^{++}$ new infections occur. The results of the tests registered at $d^{+}$ become available at $(d+1)^{-}$. If an individual is detected to be positive at time $(d+1)^{-}$, it is isolated from the population indefinitely. 

\subsection{Dorfman Adaptation}
We use two-stage Dorfman testing on a daily basis to detect infected individuals and isolate them from the population. On the first day, we apply the first stage of Dorfman test. For the first stage testing at $d^{+}$ where $d\geq 1$, only individuals from the same community are grouped together. If an individual belongs to a contaminated group at $d^{+}$, it participates in the second stage testing at $(d+1)^{+}$. Any person that belongs to an uncontaminated group at $d^{+}$ is assigned to a new group at $(d+1)^{+}$ and tested as part of the first stage testing of Dorfman testing. If a person participated in the second stage testing at $d^{+}$ and was infected, we should learn the result at $(d+1)^{-}$ and isolate that individual indefinitely. On the other hand, if that individual is not infected, it will be assigned to a new group at $(d+1)^{+}$ for a first stage Dorfman testing.

We first briefly discuss why using Dorfman testing could be more beneficial compared to non-adaptive group testing used in \cite{diggavi2021dynamic}. First, note that the use of a non-adaptive testing method, both for i.i.d.~and non-i.i.d.~priors, may result in classification errors. Hence, the assumption of independent infections argued in \cite{diggavi2021dynamic} does not hold when errors occur in decoding. Definite defective (DD) algorithm makes sure that no false-positive classification is made but it allows false-negative classifications. Moreover, since this is a dynamic problem with a time dimension, any error made at one step potentially accumulates over time. Hence, some individuals that get infected at $(d-1)^{++}$ cannot be found from the tests registered at $d^{+}$, which will be available at $(d+1)^{-}$. As a result they will spread infection until they are identified as infected. In fact, we will show with simulations how severely cumulative-error problems may affect disease spread.

In comparison, we use an adaptive testing method which has zero-error testing capacity $C_{0}=1$. Hence, we do not make classification errors. A minor problem introduced by Dorfman testing is as follows: The infection status of individuals at time $d^{+}$ who are tested with a contaminated group cannot be determined from the results available at time $(d+1)^{-}$, hence independence assumption of \cite{diggavi2021dynamic} cannot be used here. However, in order to determine an optimum Dorfman group size at $(d+1)^{+}$, we have to know the number of infections present in each community at $d^{+}$. We approximate this number with the expected number of infections $I_{G,j}$ for each contaminated first stage group $G$ with size $s$ of community $j$ as
\begin{align}
    \!\!\! \mathbb{E}\left[I_{G,j}|y_{G,j}\!=\!1\right] \!=\! \frac{\sum_{x=1}^{s} {s\choose x}p^{x}(1\!-\!p)^{s-x}x}{1\!-\!(1\!-\!p)^{s}} \!=\! \frac{sp}{1\!-\!(1\!-\!p)^{s}}
\end{align}
where $p$ is the probability of infection for each individual. Using $(1+x)^{\alpha}\approx(1+\alpha x)$, we have $\mathbb{E}[I_{G,j}|y_{G,j}=1]\approx1$. Hence each contaminated group can be assumed to contain a single defective individual. In conclusion, for an individual from community $j$, who was susceptible at $d^{+}$, the probability of being infected at $d+1$, i.e., $p_{j}^{d+1}$, can be approximated as
\begin{align}
    p_{j}^{d+1} &=1-\mathbb{P}[\text{individual is not infected during ${d^{++}}$}]\\
    &=1-(1-q_{1})^{I_{j}^{d}}(1-q_{2})^{\sum_{j\prime\neq j}I_{j\prime}^{d}} \label{iid assumption eq}
\end{align}
where $I_{j}^{d}$ is the sum of number of contaminated first stage groups and number of infected individuals in second stage tests in community $j$ at $d^{+}$. Hence, $p_{j}^{d+1}$ is used to find the optimum first stage Dorfman group size at $(d+1)^{+}$ for community $j$.

Another complication introduced by Dorfman testing is the amount of time each infected individual can potentially spread the infection. Note that, an individual who was tested together with an uncontaminated group at $(d-2)^{+}$ as part of a first stage test and who gets infected during $(d-2)^{++}$ will participate again in a first stage test at $(d-1)^{+}$, and will be individually tested at $d^{+}$. As a result, we will remove this individual from the population at $(d+1)^{-}$. Thus, this individual will spread the infection during $(d-1)^{++}$ and $d^{++}$. In the next section, we propose a novel quarantining approach and its costs in dynamic group testing to address this issue. 

\section{Dorfman Testing with Quarantine Costs} \label{sec-quarantine}
Traditional Dorfman testing finds a group size $s$ for the first stage such that the expected total number of tests over the first and second stages is minimized. In a dynamical setting where disease spreads over time, and test results are not available immediately, the time between the first and second stage test results is critical in slowing down the further spread of  infections. Hence, as a precaution, all individuals who belong to contaminated first stage groups at time $(d-1)^{+}$ can be put into quarantine from $d^{-}$ until $(d+1)^{-}$. Any individual who is quarantined yet not infected will rejoin the population at $(d+1)^{-}$. Moreover, it is not necessary to test those individuals at $(d+1)^{+}$ since they were in quarantine during $d^{++}$. Hence, with quarantine, $I_{j}^{d}$ in \eqref{iid assumption eq} decreases to number of contaminated first stage groups of tests at $d^{+}$. The decrease in $I_{j}^{d}$ decreases disease spread since $p_{j}^{d+1}$ also decreases. Moreover, as $p_{j}^{d+1}$ decreases, number of tests spent also decreases since we can mix more samples together in the first stage of Dorfman testing at $(d+1)^{+}$. However, quarantining individuals is a burden to the functioning of the society, and even more so if many of those individuals were not actually infected. As a result, in a dynamic setting with Dorfman testing, there is a trade-off between test costs and quarantine costs, as well as disease spread. In this section, we analyze effects of this trade-off on Dorfman group sizes.

Expected test cost of traditional Dorfman testing is,
\begin{align}
    \mathbb{E}[T_{j}]=\frac{N_{j}}{s}+N_{j} (1-(1-p)^s)
    \label{dorfman-expected-cost}
\end{align}
where $s$ is the first stage Dorfman group size, $N_{j}$ is the size of community $j$, and $p$ is the infection prevalence rate.
We model the quarantine cost as an exponential function of the number of people who were not infected but put into quarantine,
\begin{align}
    C_{Q,j}(x)=a^{x}
    \label{C-q-cost}
\end{align}
where $a>1$ is a design choice and $x$ is the number of people in community $j$ who are quarantined unnecessarily. The quarantine cost of each community is independent of others. 

For analytical tractability purposes, we make a mild assumption that the first stage groups in each community $j$ are also independent. Although this undermines the exponential nature of the cost introduced in \eqref{C-q-cost}, without this assumption, optimum Dorfman size would depend on community size, which is not desirable for finding cost per person. However, the assumption still captures exponential behavior and the trade-off between quarantine versus test costs which is our main point here. Based on this assumption, the expected quarantine cost is,
\begin{align}
    \!\!\mathbb{E}[C_{Q,j}] \!=& \frac{N_{j}}{s}\left( \sum_{i=1}^{s-1} {s\choose i} p^{s-i}(1-p)^{i}a^{i} \right) \\
    =&\frac{N_{j}}{s} \left( p^s\sum_{i=0}^{s} {s\choose i} \left(\frac{a\!-\!ap}{p}\right)^{i} \!\!-\! (1\!-\!p)^s a^s \!-\!p^s \right) \\
    =&\frac{N_{j}}{s}\left(p^s  \left(1+\frac{a\!-\!ap}{p}\right)^{s} \!\!-\!(1\!-\!p)^s a^s \!-\!p^s\right) \\
    =&\frac{N_{j}}{s} \left((a-ap+p)^s-(a-ap)^s-p^s \right) \label{expected-quarantine-cost}
\end{align}

\begin{figure}[t]
   \centerline{\includegraphics[width=.93\columnwidth]{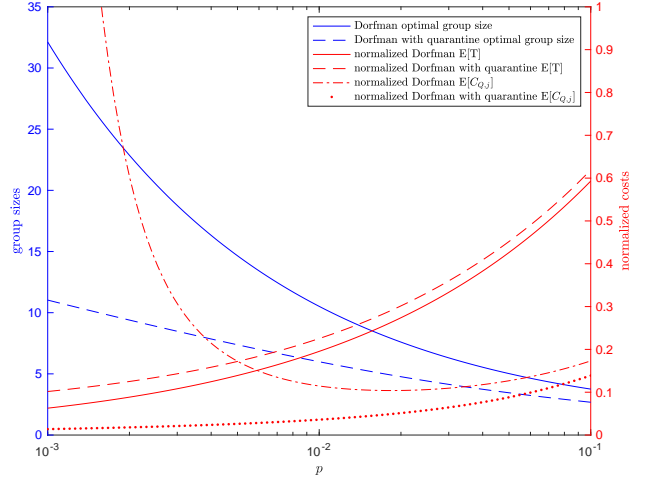}}
   \caption{Test cost optimization versus test and quarantine cost optimization.}
   \label{dorf-iso-1.3-2}
   \vspace*{-0.2cm}
\end{figure}

Next, instead of minimizing the test cost in \eqref{dorfman-expected-cost} alone, we minimize the weighted average of the test cost in \eqref{dorfman-expected-cost} and the quarantine cost in \eqref{expected-quarantine-cost}, where $\alpha$ is a design choice that determines the relative importance of test and quarantine costs, 
\begin{align}
    \frac{\mathbb{E}[T_{j}]\!+\!\alpha\mathbb{E}[C_{Q,j}]}{N_{j}} \!=&\frac{1}{s}\!+\!(1\!-\!(1\!-\!p)^s) \nonumber \\
    &+\!\frac{\alpha}{s} \!\left((a\!-\!ap\!+\!p)^s\!-\!(a\!-\!ap)^s\!-\!p^s \right) 
    \label{quarantine-dorfman-expected cost}
\end{align}

\figref{dorf-iso-1.3-2} shows a comparison between the normalized Dorfman test cost alone, and weighted Dorfman test and quarantine costs, for $(a,\alpha)=(1.3,2)$. We plot optimal group sizes for each $p$ that are logarithmically spaced between $[10^{-3}, 10^{-1}]$. On the left $y$-axis, we show optimal group sizes obtained by optimizing \eqref{dorfman-expected-cost} and \eqref{quarantine-dorfman-expected cost}, respectively. On the right $y$-axis, we show optimized normalized test and quarantine costs. As each unnecessarily quarantined individual has an exponential effect, we see that normalized quarantine cost of optimized \eqref{dorfman-expected-cost} is greater by orders of magnitude than optimized \eqref{quarantine-dorfman-expected cost}, whereas normalized test costs are still of the same order. Hence, in settings with moderate $a$ and small to moderate $p$, optimizing \eqref{quarantine-dorfman-expected cost} instead of \eqref{dorfman-expected-cost} decreases quarantine cost substantially without effecting test cost as much.

\begin{figure}[t]
   \centerline{\includegraphics[width=\columnwidth]{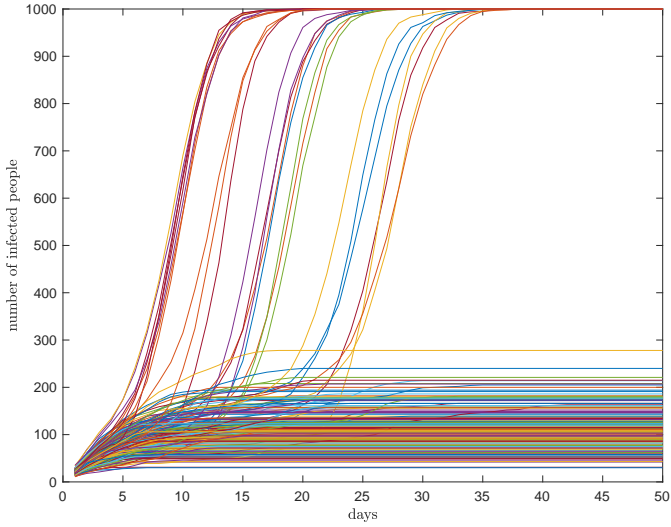}}
   \caption{CCA method, cumulative infection trajectories.}
   \label{diggavi-realistic-cca-inf-trajectories}
\end{figure}

\begin{figure}[t]
   \centerline{\includegraphics[width=.97\columnwidth]{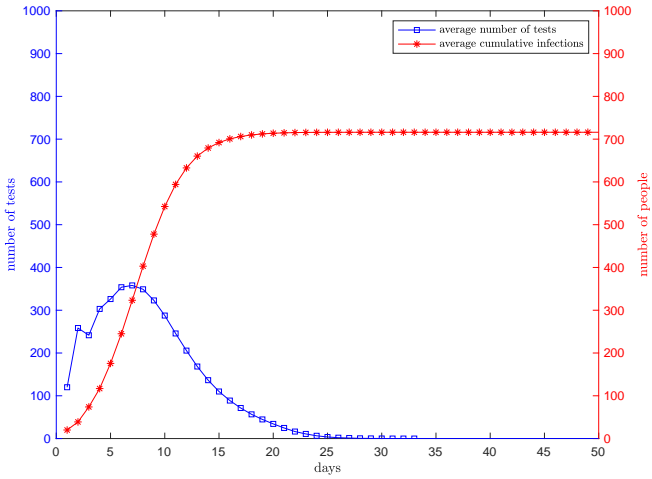}}
   \caption{Dynamic Dorfman testing without quarantine.}
   \label{regular-dorfman-sbm}
\end{figure}

\subsection{Experimental Results}
We consider the system model introduced in Section \ref{SIRSBM-section} with parameters SBM $(1000, 50, 0.012, 0.0004, 0.02)$ and $r=0.1$. First, we experimentally show how a non-adaptive method like CCA cannot overcome the cumulative error introduced by dynamic settings if it uses comparable number of tests as dynamic Dorfman testing. Normally, on day $d$ with population size $N_{d}$ for CCA algorithm, we would have to choose number of tests as $4e(1+\delta)\mu\ln N_{d}$. However, as mentioned in Appendix of \cite{diggavi2021dynamic}, the number of tests that will be used during initial days will be at least as large as $N$ and group testing will offer no advantage. Hence, for the purpose of comparing adaptive Dorfman algorithm with CCA, here we choose $T_{d}=1.6e\mu\ln N_{d}$, which is on average more than what Dorfman testing would have used on a daily basis with these system parameters. \figref{diggavi-realistic-cca-inf-trajectories} shows total number of infected people during a $50$ day testing period for $200$ different trajectories. Here, approximately $13\%$ of these trajectories end up with disease explosion due to cumulative errors. 

\begin{figure}[t]
   \centerline{\includegraphics[width=.97\columnwidth]{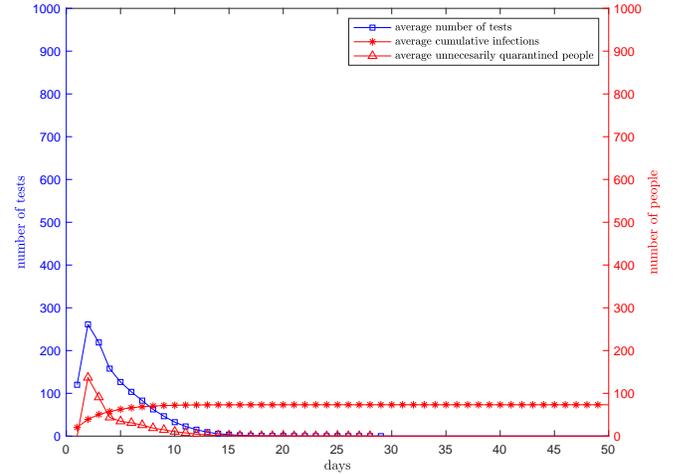}}
   \caption{Dynamic Dorfman testing with quarantine without quarantine cost.}
   \label{quarantine-dorfman-sbm-no-cost}
\end{figure}

\begin{figure}[t]
   \centerline{\includegraphics[width=.97\columnwidth]{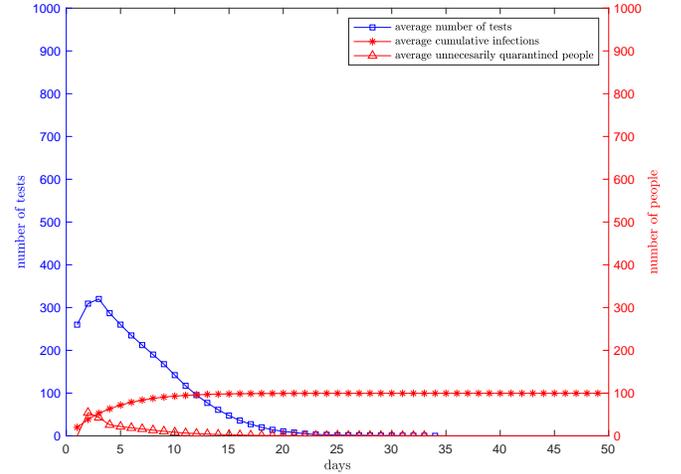}}
   \caption{Dynamic Dorfman testing with quarantine and its cost.}
   \label{quarantine-dorfman-sbm-with-cost}
\end{figure}

Next, we examine how Dorfman testing performs for the same SBM setting for $1000$ different trajectories. In \figref{regular-dorfman-sbm}, experimental results for Dorfman testing are displayed where we do not quarantine people who belong to contaminated groups and isolation is only applicable for individuals who test positive in the second stage. Without a quarantine procedure, $71\%$ of population contracts the disease, which is undesirable.

In \figref{quarantine-dorfman-sbm-no-cost}, we show how these results change if we apply quarantine to contaminated groups based on the first stage results and individuals who are not infected yet quarantined join the population once their individual test results are available next day. Note that disease progression is decreased since an individual who gets infected during $(d-2)^{++}$ can spread infection during $(d-1)^{++}$ instead of during both $(d-1)^{++}$ and $d^{++}$. As a result, number of tests spent decreases since $p_{j}^{d+1}$ decreases. Although the results are promising since we spend a small amount of tests and only $7\%$ of the population contracts the disease, unnecessary quarantine numbers are quite large which is undesirable. 

Finally, when we use the objective in \eqref{quarantine-dorfman-expected cost} instead of that in \eqref{dorfman-expected-cost}, and and choose $(a,\alpha)=(1.5,2)$, we obtain the results shown in \figref{quarantine-dorfman-sbm-with-cost}. We observe that we can decrease quarantine costs substantially while increasing test costs. Also note that, total disease spread has increased from $7\%$ to $10\%$ since fewer uninfected individuals spend one day in quarantine during outbreak which results in vulnerability to contract the disease. Thus, \eqref{quarantine-dorfman-expected cost} introduces a useful tool to trade-off quarantine costs, test costs, and disease spread. 

\section{Optimizing Group Sizes for Dynamic Model with i.i.d.~Infections over Time}
In static Dorfman testing, the expected number of tests is minimized for a single day. For a dynamical model, on each day, group size for the first stage testing directly affects the number of people participating in the first stage of testing on the next day. Hence, optimizing the test cost for a single day may not be optimal for the entire horizon of testing. In this section, we investigate this problem for a tractable i.i.d.~setting.

\subsection{System Model}
Here, we consider the following simplified i.i.d.~infection model: Each day, each individual that has not been infected so far can get infected i.i.d.~with probability $p$. Any individual that has been infected is isolated indefinitely once it is classified as infected via Dorfman testing. 

\subsection{Test Costs}
Let $N_{d}$ denote the number of people entering the first stage Dorfman testing on day $d$. Together with the second stage testing of these individuals, the expected number of total first stage tests on day $d$ and second stage (individual) tests that will be registered on day $d+1$ is,
\begin{align}
    \mathbb{E}\left[T_{d}|N_{d}\right]=\frac{N_{d}}{s_{d}}+N_d(1-(1-p)^{s_{d}}) \label{single day dorfman}
\end{align}
and therefore,
\begin{align}
    \mathbb{E}\left[T_{d}\right]=\frac{\mathbb{E}\left[N_{d}\right]}{s_{d}}+\mathbb{E}\left[N_{d}\right](1-(1-p)^{s_{d}})
\end{align}
where $s_{d}$ denotes the first stage group size on day $d$. $N_{d}$ is a random variable which depends on $N_{1}^{d-1}$ and $s_{1}^{d-1}$ as well as $p$, where $s_{1}^{d-1}$ denotes $[s_{1},\ldots,s_{d-1}]$ and similarly for $N_{1}^{d-1}$. To express $N_{d}$ in terms of these variables, let $h_{d-1}$ be the number of uncontaminated groups found from results of day $d-1$ and $g_{d-2}$ be the number of contaminated groups found from results of day $d-2$. Note that $(h_{d}+g_{d})s_{d}=N_{d}$. Also, let $h_{d-2,k}$ be the number of negative individuals of a contaminated group $k$, which were tested together with some infected individuals on day $d-2$ as part of Dorfman first stage testing and let $h_{d-2,k}^{-}$ denote the individuals who do not catch the infection before the second stage test on day $d-1$. We can express $N_{d}$ in terms of these variables as,
\begin{align}
    N_{d}=h_{d-1}s_{d-1}+\sum_{k=1}^{g_{d-2}}h_{d-2,k}^{-}
    \label{conv_day_equation}
\end{align}
Then, we find expectation of each of these variables as,
\begin{align}
    \mathbb{E}\left[h_{d-1}|N_{d-1}\right]&=\frac{N_{d-1}}{s_{d-1}}(1-p)^{s_{d-1}}\\
    \mathbb{E}\left[h_{d-1}\right]&=\frac{\mathbb{E}\left[N_{d-1}\right]}{s_{d-1}}(1-p)^{s_{d-1}}
\end{align}
\begin{align}
    \mathbb{E}\left[g_{d-2}|N_{d-2}\right]&=\frac{N_{d-2}}{s_{d-2}}(1-(1-p)^{s_{d-2}})\\
    \mathbb{E}\left[g_{d-2}\right]&=\frac{\mathbb{E}\left[N_{d-2}\right]}{s_{d-2}}(1-(1-p)^{s_{d-2}})
\end{align}
\begin{align}
    \mathbb{E}\left[h_{d-2,k}\right]&=s_{d-2}\left(1-\frac{p}{1-(1-p)^{s_{d-2}}}\right) \\
    \mathbb{E}\left[h_{d-2,k}^{-}|h_{d-2,k}\right]&=h_{d-2,k}(1-p) \\
    \mathbb{E}\left[h_{d-2,k}^{-}\right]&=s_{d-2}(1-p)\left(1-\frac{p}{1-(1-p)^{s_{d-2}}}\right)
\end{align}

Hence, by taking expectation of \eqref{conv_day_equation}, we obtain,
\begin{align}
     \mathbb{E}\left[N_{d}\right]&= \mathbb{E}\left[h_{d-1}\right]s_{d-1}+\mathbb{E}\left[g_{d-2}\right]\mathbb{E}\left[h_{d-2,k}^{-}\right] \\
     &= \mathbb{E}\left[N_{d-1}\right](1-p)^{s_{d-1}}  \nonumber \\
     &\quad + \mathbb{E}\left[N_{d-2}\right](1-(1-p)^{s_{d-2}}-p)(1-p) \label{exp_conv_day_equation}
\end{align}
Assuming that testing will take place between day $1$ and day $t$, where individual testing stage of day $t$ will happen on day $t+1$, the total expected number of tests can be expressed as,
\begin{align}
    \mathbb{E}\left[T_{tot}\right]=&\sum_{d=1}^{t}\mathbb{E}\left[T_{d}\right] \\
    =&\sum_{d=1}^{t}\frac{\mathbb{E}\left[N_{d}\right]}{s_{d}}+\mathbb{E}\left[N_{d}\right](1-(1-p)^{s_{d}}) \label{tot_cost_equation}
\end{align}
This equation, in turn, can be solely expressed in terms of $\mathbb{E}\left[N_{1}\right]=N_{1}$ and $p$ by using \eqref{exp_conv_day_equation} iteratively. Thus,
\begin{align}
    \mathbb{E}\left[T_{tot}\right]=f(N_{1}, p, s_{1}^{t}) \label{init-dorfman-iid-opt}
\end{align}

\begin{figure}[t]
	\centerline{\includegraphics[width=.95\columnwidth]{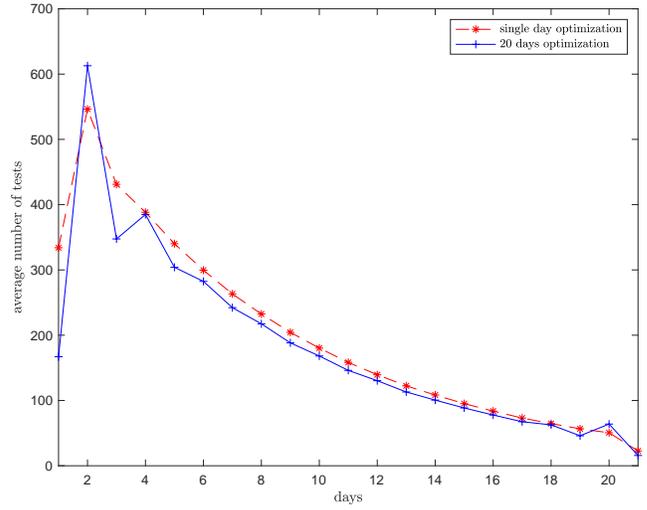}}
	\caption{Optimization of a single day versus optimization of the entire horizon.}
	\label{indep-fullvsone}
	\vspace*{-0.2cm}
\end{figure}

After $\mathbb{E}\left[T_{tot}\right]$ is found, we can optimize this cost over $s_{1}^{t}$. Let $s_{d}^{*}$ denote the optimum value for day $d$. By inspecting \eqref{exp_conv_day_equation} and \eqref{tot_cost_equation}, we see that $s_{t}^{*}$ does not depend on $s_{1}^{t-1}$ even though $\mathbb{E}\left[N_{t}\right]$ depends on $s_{1}^{t-1}$ since $\mathbb{E}\left[N_{t}\right]$ appears as normalization term. As a result, $s_{t}^{*}$ is uniquely determined by the value of $p$. Once this is done, $s_{t-1}^{*}$ can be found using $s_{t}^{*}$ and $p$. This can be iteratively done for all $s_{d}^{*}$ going backwards from $s_{t}^{*}$ to $s_{1}^{*}$. Once we find $s_{1}^{*}$, we proceed with the first stage Dorfman testing for day $d=1$. After that, on day $d=2$, we obtain the first stage results of day $d=1$ and we have to optimize the cost in \eqref{tot_cost_equation} for $s_{2}^{t}$ where $h_{1}$ (and $g_{1}$) is no longer a random variable. We observe that $h_{1}$ impacts $\mathbb{E}\left[N_{d}\right]$ for $d \geq 2$. But this results in the same optimum values for $s_{2}^{t}$ since each $s_{d}^{*}$ only depends on $s_{d+1}^{t}$ and $h_{1}$ can only add a normalization factor for $\mathbb{E}\left[N_{d}\right]$ for $d \geq 2$. The same argument can be done for realization of $h_{d-2,k}^{-}$ on day $d$ for each $k$. Hence, no matter what realization comes out for $h_{d-1}$ and $h_{d-2,k}^{-}$, $s_{d}^{*}$ will be the same as initial optimum value found from \eqref{init-dorfman-iid-opt}, which means we do not have to worry about realizations of $h_{1},\ldots,h_{t-1}$ and $h_{1,k}^{-},\ldots,h_{t-2,k}^{-}$.

\begin{figure}[t]
	\centerline{\includegraphics[width=.97\columnwidth]{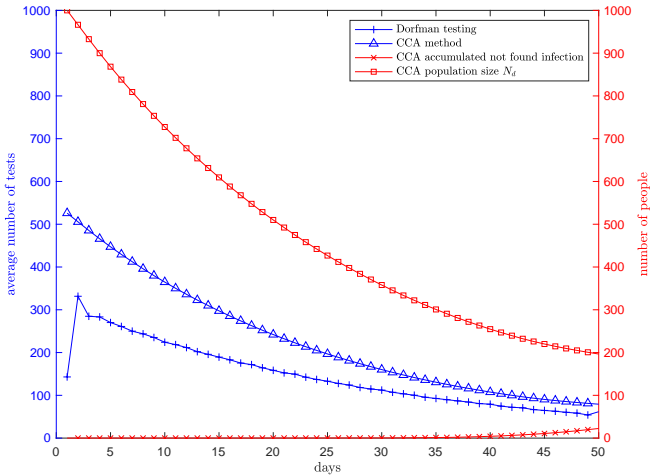}}
	\caption{CCA versus Dorfman testing for i.i.d.~infection model with $T_{d}=0.8epN_{d}\log(N_{d})$.}
	\label{ccavsdorf-iid}
	\vspace*{-0.2cm}
\end{figure}

\subsection{Experimental Results}
In \figref{indep-fullvsone}, we compare the performance of optimizing Dorfman group sizes on a daily basis using \eqref{single day dorfman} with optimizing \eqref{init-dorfman-iid-opt}. Here, we choose $N=1000$, $t=20$, $p=0.12$. The curves show that by optimizing \eqref{init-dorfman-iid-opt} we can get some benefits in terms of number of tests spent. Although optimum values of \eqref{init-dorfman-iid-opt} are different from \eqref{single day dorfman}, rounding the group sizes to integer values removes some of the benefits of optimizing \eqref{init-dorfman-iid-opt}.

\begin{figure}[t]
	\centerline{\includegraphics[width=.97\columnwidth]{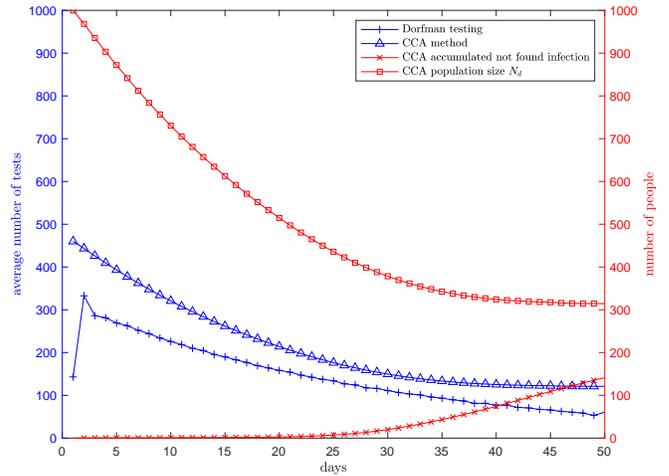}}
	\caption{CCA versus Dorfman testing for i.i.d.~infection model with $T_{d}=0.7epN_{d}\log(N_{d})$.}
	\label{ccavsdorf-iid-explode}
	\vspace*{-0.2cm}
\end{figure}

In \figref{ccavsdorf-iid} and \figref{ccavsdorf-iid-explode}, we compare the performance of Dorfman testing using \eqref{init-dorfman-iid-opt} with dynamic non-adaptive group testing (CCA) using two different number of tests per day, $T_{d}$, values. The infection model is as stated above, i.e., each day infections occur i.i.d.~with $N=1000$, $t=50$, $p=0.035$. Here, we also show daily population size $N_{d}$, as well as the number of people that have not been detected for more than $2$ days for the CCA algorithm (note that Dorfman testing requires at most $2$ days to find infected individuals). For CCA, if we chose the number of tests as $T_{d}=0.8epN_{d}\ln N_{d}$, then, Dorfman testing outperforms CCA. If we decrease $0.8$ factor to $0.7$ as in \figref{ccavsdorf-iid-explode}, the number of undetected people explodes for the CCA algorithm.

\section{Conclusion}
In this paper, we introduced an adaptation of Dorfman group testing for dynamical settings with disease spread over time. We investigated the benefits as well as complications of the proposal adaptation, and compared its performance to that of a state-of-the-art non-adaptive testing method. By introducing the quarantine concept and its costs, we showed how optimal Dorfman group sizes can change, and investigated the trade-off between the test cost, quarantine cost, and disease progression. This theoretical quarantine modeling is similar to practically implemented close-contact tracing in communities during the ongoing covid-19 pandemic, where accurate tests like PCR may result in delays, during which time the individuals may be presumed potentially infected, and may be asked to be quarantined. We also showed that optimum dynamic Dorfman group sizes can differ from the static case in i.i.d.~settings. 

\bibliographystyle{unsrt}
\bibliography{lib}

\end{document}